# TAUP-2647
November 19, 2000# Momentum Reconstruction and Triggering in the ATLAS Detector

Gideon Dror[1+], Erez Etzion[2*]

1. Department of Computer Science, The Academic College of Tel-Aviv-Yaffo, Tel-Aviv 64044, Israel.
2. School of Physics and Astronomy, Raymond and Beverly Sackler Faculty of Exact Sciences, Tel-Aviv University, Tel-Aviv 69978, Israel.**Abstract.** A neural network solution for a complicated experimental High Energy Physics problem is described. The method is used to reconstruct the momentum and charge of muons produced in collisions of particle in the ATLAS detector. The information used for the reconstruction is limited to the output of the outer layer of the detector, after the muons went through strong and inhomogeneous magnetic field that have bent their trajectory. It is demonstrated that neural network solution is efficient in performing this task. It is shown that this mechanism can be efficient in rapid classification as required in triggering systems of the future particle accelerators. The parallel processing nature of the network makes it relevant for hardware realization in the ATLAS triggering system.## INTRODUCTION

The Large Hadron Collider (LHC) currently under construction is the largest particle accelerator ever built. It is scheduled to start operating in 5 years. The design luminosity is so high that a rate of 1 GHz of events is expected in its detectors. In such an environment it is important to distinguish between the relevant physics collision within a time frame of a few nano seconds. The goal of the trigger is to select 100 out of 1 billion events per second. One of the ways to select interesting events is to look for the signature of high transverse momentum ($P_T$) muons. Going through the ATLAS magnetic field the muons momentum could be derived from their curvature. The complicated inhomogeneous magnetic field in the forward area of the ATLAS detector makes it a very complicated problem to solve. However it is shown that a neural network implementation can be used for a fast and efficient triggering system. A standard feed forward network is trained to learn the characteristics of the detectors output, and to solve their complicated dynamics. The network successfully solves the inverse problem described below, and can efficiently be implemented in a hardware based triggering system

## EXPERIMENTAL SETTING

The goal is to use the momentum of muons produced in the interaction point of the ATLAS detector as a triggering signal. Muons with low angles with respect to the beam line cross the tracking device, shielding and calorimetry before they reach the Thin Gap Chambers (TGC) [1] located at the outer parts of the detector, 12-14 meters away from the interaction point. The muons emerging in the end-caps

**FIGURE 1.** A side view of one ATLAS's quadrant. In black the muons trajectories. In brown the three TGCs layers.

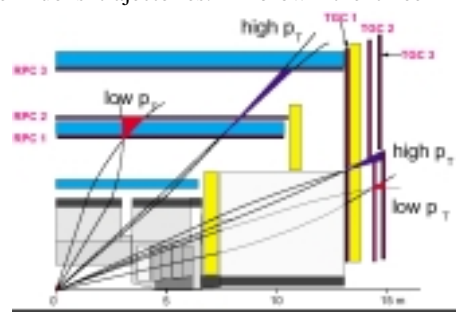

---

[∴] Presented at ACAT 2000 FermiLab, Chicago October 2000
[+] Gideon@server.mta.ac.il
[*] Erez@lep.tau.ac.il

undergo stochastic electric Coulomb scattering and their tracks are bent by highly inhomogeneous magnetic fields. The target is to deduce from the TGC hits the transverse momentum, $\vec{P}_T$, with which the muon was created. As shown in Fig.1 the TGC coverage is limited to low polar angles corresponding to pseudorapidity $\eta > 1.05$.

## THE NEURAL NETWORK

A standard feed forward was used to learn the mapping between the final muon track as detected by the TGC hits and its initial charge and momentum. Training was done with Lvenberg-Marquardt method on simulated events generated with the ATLAS simulation program, DICE[2],. The events studied were single muons in the area of $\eta > 1.05$ and $1 < \vec{P}_T < 50 \, \text{GeV}$.

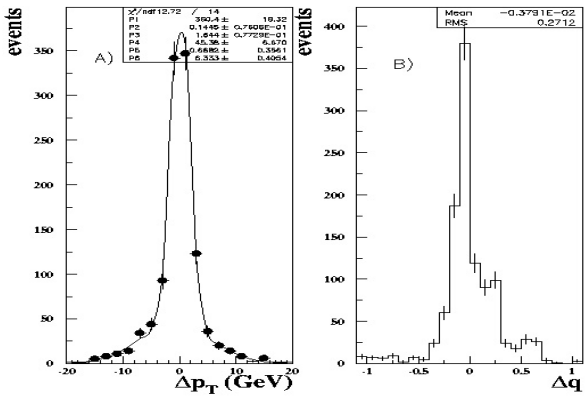

**FIGURE 2.** Differences between calculated and generated momentum, left - $\vec{P}_T$, right - charge.

The NN is fed with 4 inputs, the slopes and intercepts of the projections of muon's tracks on *xz* an *yz* planes, calculated from LSQ linear fit. This preprocessing is required since while some of the hits in the raw data are missing, there are multiple hits in regions where TGC plates overlap. Earlier studies[3] used the Hough transform for preprocessing, however it turned out that the LSQ fit is sufficient. Four output linear neurons provide the value of the muon initial $\vec{P}_T$, initial direction and its electric charge, q. Several NN architectures were examined. The chosen NN contains two hidden layers with seven sigmoidal transfer function each. Each network was trained using Levenberg-Marquardt method with 2500 events where the minimal number of events sufficient for reasonable solution is about 1400 events. Typically the training lasted several thousands epochs. Upon completion of the training phase each network was examined with a test set of 1829 events.

## RESULTS

The difference between the NN estimates and the real $\vec{P}_T$ and charge as determined for the test set are shown in Fig. 2. The width of the two result among other from the real stochasticity distribution of the data due to the interaction point widths as well as the Coulomb scattering.

The relative error on $\vec{P}_T$ as a function of $\eta$ is shown in Fig. 3. The poor resolution for $\eta \approx 1.7$ results from the highly inhomogeneous magnetic field in the region between the end-cap and the barrel ATLAS toroid coils.

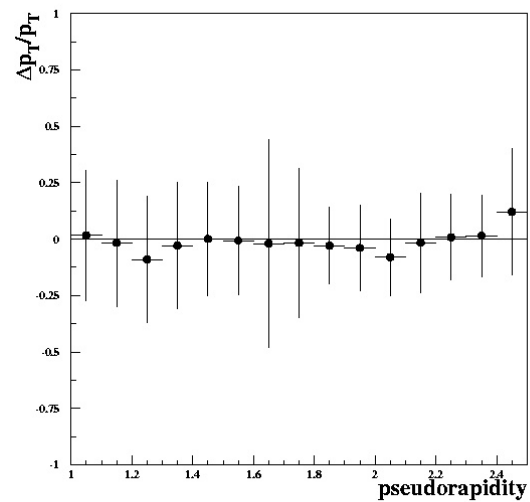

**FIGURE 3.** Mean relative error on $\vec{P}_T$ as a function of $\eta$. The error bars are the RMS of Gaussian fit of $\Delta P_T / P_T$.

The identification of the charge is more robust due to its discrete nature. In 98.5% of the events the charge was correctly identified, where naturally most of the misclassification occurs at high momentum.

## CLASSIFIACTION NETWORK

The speed of the network decision is crucial in triggering network. In our case the decision is simply based on the transverse momentum of the muon being bigger or smaller than a certain value (6 GeV). Only events with transverse momentum larger than 6 are stored for further analysis. The network described above yields about 94% correct classifications. As can be expected the misclassifications occur mainly near the threshold value.

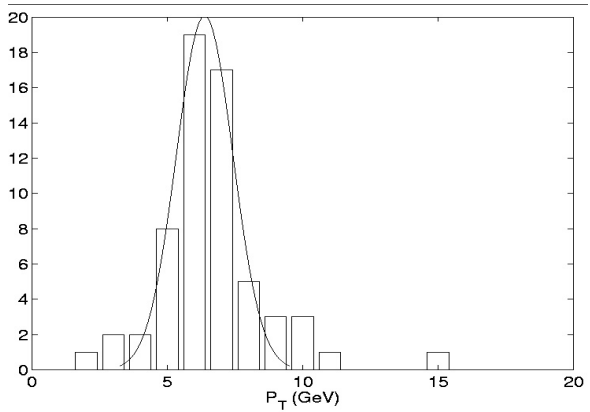

**FIGURE 4**. A Gaussin fit to the real $\vec{P}_T$ of the events which were misclassified by the trakcing and selecting NN. The mean and RMS of the fit are 6.4 and 1.0 GeV respectively.

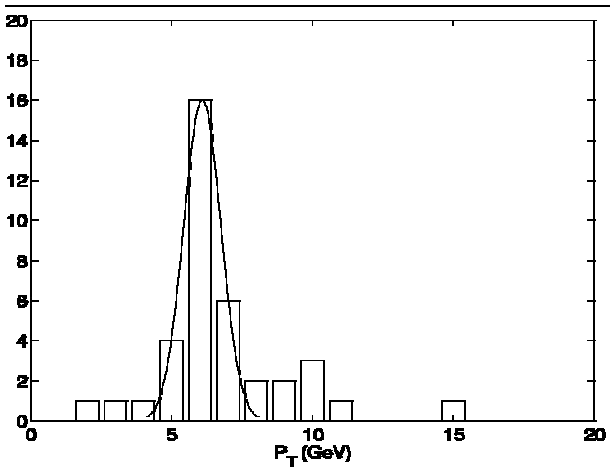

**FIGURE 5.** A Gaussin fit to the real $\vec{P}_T$ of the events that were misclassified by selection dedicated NN. The mean and RMS of the fit are 6.1 and 0.7 GeV respectively.

A more appropriate approach for triggering purposes is to directly train the network on the classification problem only. We used for this purpose a network with similar architecture, replacing the linear output neurons with a sigmoidal one.

We find that 96% of the events are correctly classified where again most misclassification occur near the threshold of transverse momentum around 6 GeV.

The $\vec{P}_T$ distributions of the misclassified events in the two methods are shown in Figs. 4 and 5. It turned out that most error are common to both networks and stem from misleading events produced in rare large-angle scattering. However, note that the width of the dedicated classification network is considerably narrower.

## DISCUSSION

A feed forward NN capable of estimating the charge and momentum of muons in the ATLAS particle facility currently under construction is presented. With relatively simple architecture we were able to solve a complicated inverse problem. A similar NN can very efficiently be used in classification problem necessary for triggering purposes. Due to the parallel nature of the network, it opens a way for hardware realization of this problem. A. Chorti et al., presented in the current workshop, a successful test of a hardware implementation of this exact architecture [4].

## ACKNOWLEDGMENTS

We wish to acknowledge the fruitful collaboration and the stimulating discussions with our colleagues Halina Abramowicz and David Horn from Tel-Aviv University. The research was partly supported by the Israeli Science Foundation.

## REFERENCES

1. G. Bella et al., *Nucl. Inst. Meth* **A252**, 503 (1986); S. Dado et al., *Nucl. Inst. Meth* **A252**, 511 (1986); G. Mikenberg et al., *Nucl. Inst. Meth* **A265**, 23 (1988).

2. DICE Manual, *ATLAS note* **SOFT-NO-10** (1994).

3. G. Dror et al., *AIHENP99 proceedings*, edited by G. Athanasiu, D. Perret-Gallix, Elsevier North-Holland Editors, (1999).

4. A. Chorti, B. Granado, B. Denby, et. al. "An electronic system for simulation of neural networks with a micro-second real time constraint" ACAT 2000.